\documentclass[9pt]{article}

\usepackage{url}
\usepackage{caption}
\usepackage[usenames,dvipsnames]{xcolor}
\usepackage{ragged2e}
\usepackage[cmex10]{amsmath}
\usepackage{float}
\usepackage{graphicx}
\usepackage{amssymb,amsfonts}
\usepackage{fullpage}
\usepackage{times}
\usepackage{overpic}

\graphicspath{{figures/}}
\DeclareGraphicsExtensions{.pdf,.png,.jpg}

\newcommand{\reffig}[1]{Figure\,\ref{#1}}
\newcommand{\refeq}[1]{equation\,(\ref{#1})}
\newcommand{\reftab}[1]{Table\,\ref{#1}}

\makeatletter
\renewcommand\@biblabel[1]{#1 }

\def\@maketitle{%
  \newpage\spacing{1}\setlength{\parskip}{10pt}%
    \begin{center}%
	{\large\bfseries\noindent\sloppy \textsf{\@title} \par}%
    {\noindent\sloppy \@author}%
	\end{center}%
}

\renewcommand{\section}{\@startsection {section}{1}{0pt}%
    {-6pt}{1pt}%
    {\reset@font \large \bfseries}%
    }
\renewcommand{\subsection}{\@startsection {subsection}{2}{0pt}%
    {-6pt}{1pt}%
    {\reset@font \normalsize \bfseries}%
    }
\makeatother

\newcommand{\spacing}[1]{\renewcommand{\baselinestretch}{#1}\large\normalsize}

\AtBeginDocument{}

\begin{document}

% Title must be 150 characters or less
\begin{flushleft}
	{\Large
		\textbf{Hybrid Epidemics - A Case Study on Computer Worm Conficker}
	}
    \newline
	% Insert Author names, affiliations and corresponding author email.
	\\
	Changwang Zhang$^{1,2,\ast}$, 
	Shi Zhou$^{1,\ast}$,
    and Benjamin M. Chain$^{3}$
	\\
    \bigskip
	\bf{1} Department of Computer Science, University College London, UK
	\\
	\bf{2} Security Science Doctoral Research Training Centre, University College London, UK
	\\
	\bf{3} Division of Infection and Immunity, University College London, UK
	\\
    \bigskip
	$\ast$ E-mail: changwang.zhang.10@ucl.ac.uk (CZ); s.zhou@ucl.ac.uk (SZ)
\end{flushleft}

\section*{Abstract} % (Not exceed 300 words)
Conficker is a computer worm that erupted on the Internet in 2008. It is unique in combining three different spreading strategies: local probing, neighbourhood probing, and global probing. We propose a mathematical model that combines three modes of spreading, local, neighbourhood and global to capture the worm's spreading behaviour. The parameters of the model are inferred directly from network data obtained during the first day of the Conifcker epidemic. The model is then used to explore the trade-off between spreading modes in determining the worm's effectiveness. Our results show that the Conficker epidemic is an example of a critically hybrid epidemic, in which the different modes of spreading in isolation do not lead to successful epidemics. Such hybrid spreading strategies may be used beneficially to provide the most effective strategies for promulgating information across a large population. When used maliciously, however, they can present a dangerous challenge to current internet security protocols. 

\section*{Introduction}

Epidemic spreading phenomena exist in a wide range of domains \cite{House_2012,Rock_2014}. Well-known examples include disease spreading \cite{R_Zheng_2014_3,R_Nie_2014,R_Li_2014}, computer worm proliferation \cite{Zou_2006,Shin_2012,Yu_2015}, and information propagation \cite{Ren_2014,Chen_2014,Sahneh_2014}. Modelling and understanding of such phenomena can have important practical values to predict and control real world epidemics \cite{R_Zheng_2014_3,R_Nie_2014,R_Li_2014,R_Zhang_2014_2,R_Zhang_2014,Moore_2003,Shannon_Moore_2004}.

%Modelling and understanding of such phenomena can have important practical values. For example, epidemic studies \cite{R_Zhang_2014_2, R_Zheng_2014_3,R_Nie_2014,R_Li_2014} were proposed to predict and control the transmission of the brucellosis, which is a disease that infects both humans and animals. Epidemic models were also used to help determine the original source of Avian influenza \cite{R_Zhang_2014}.

Some typical spreading mechanisms have been extensively studied, such as the {fully-mixed} spreading model and the {network} spreading model. Many epidemics are {\em hybrid} as they spread via two or more different mechanisms simultaneously.
Previous work on hybrid epidemics has focused on what we call the {\em non-critically} hybrid epidemic, where at least one of the spreading mechanisms alone is able to cause an epidemic outbreak, and a mixture of mechanisms brings no advantage. 

We are interested in the {\em critically} hybrid epidemic, where each  spreading mechanism alone is unable to cause any significant spreading whereas the mixture of such mechanisms leads to a huge epidemic outbreak. 
Recently we proposed a model that explains the behaviour of critically hybrid epidemics, which incorporates two spreading mechanisms in the setting of a metopopulation \cite{Zhang_hm_2014}. We demonstrated that it is indeed possible to have 
a highly contagious epidemic by mixing simple, ineffective spreading mechanisms. The properties of such epidemics are critically determined by the ratio at which the different spreading mechanisms are mixed, and usually there is an optimal ratio that leads to a maximal outbreak size. 

In this paper we present a detailed analysis of a {\em real} hybrid epidemic -- the Internet worm Conficker, which erupted on the Internet in 2008 and infected millions of computers . 
The worm is a hybrid epidemic as the code analysis \cite{Eric_2010} has revealed the worm applied three distinct spreading mechanisms: (1) global random spreading, (2) local network spreading, and (3) neighbourhood spreading. 
It is a critically hybrid epidemic because the first and second spreading mechanisms are highly ineffective if used alone, and the third mechanism, as we will show later, is most effective when mixed with the other two.   

We introduce a mathematical model to describe the spreading behaviour of Conficker. Our study was based on measurement data provided by Center for Applied Internet Data Analysis (CAIDA)'s Network Telescope project~\cite{CAIDA_2008_aft, CAIDA_2008_bfr}, which monitors Internet traffic anomalies. 
We proposed algorithms to extract Conficker--related features from the CAIDA data. Then we infer the values of our model's parameters that characterise the worm. 

We evaluated our inference results by comparing theoretical predictions with the actual measurement results. Our predictions closely reproduced the outbreak process of Conficker. 
We then explored possible spreading scenarios based on simulations using different values of parameters. 
One of the interesting results was that we showed the worm could spread faster, reach a larger outbreak size or survive for longer time by just revising the ratios at which the worm allocated its time on each of the spreading mechanisms (while keeping everything else the same), which can be easily achieved by changing a few lines in its coding.

This paper's contributions are two fold. 
Firstly, we present the first study on a {real-life} critically hybrid epidemic, where the epidemic's parameter values are inferred from measurement data.
Secondly, we analyse the complex interactions among Conficker's three spreading mechanisms, and show that the worm can be more contagious if it mixes its three spreading mechanisms in an optimal way.
%
%Thirdly, our work highlights the practical importance of studying the critically hybrid epidemics, for example to contain malicious spreading (e.g.~Internet worms) as well as to facilitate beneficial spreading (e.g.~advertisement).

\section*{Background}

\subsection*{Epidemic spreading mechanisms}

A number of epidemic spreading mechanisms have been extensively studied~\cite{Newman_Book_2010,Keeling_Eames_2005}. 
For example, in the {\em fully-mixed} spreading models \cite{Newman_Book_2010,Anderson_1991}, a node is connected to all other nodes in a population, thus an epidemic can potentially spread between any two nodes according to a probability. %Example include a flu outbreak in a city. 
Whereas in the {\em network} spreading models \cite{House_2012,Rock_2014, Newman_Book_2010, Pastor-Satorras_Vespignani_2001}, nodes are connected to their neighbours via a network structure, therefore an epidemic can only spread along the connections among nodes. Recent network-based models considered additional physical properties such as location-specific contact patterns \cite{R_Wang_2013, R_Wang_2013_2}, human mobility patterns \cite{Balcan_2009b, Wang_2009, Balcan_2011, Meloni_2011} and spatial effects \cite{R_Sun_2007,R_Sun_2008,R_Sun_2010,R_Sun_2012}.

\subsection*{Hybrid epidemics}

Many epidemics are {\em hybrid} in the sense that they spread via two or more spreading mechanisms simultaneously.
A hybrid epidemic can use fully-mixed spreading and network spreading, or use  fully-mixed spreading but at two or more different levels, e.g. at the {\em global} level covering the whole population or at the {\em local} level consisting of only a part of the population.

There are many real examples. Mobile phone viruses can spread via Bluetooth communication with any nearby devises (local, fully-mixed spreading) and Multimedia Messaging Service with remote contacts (global, network spreading) \cite{Wang_2009}.
A computer that is infected by the worm {Red Code II} spends 1/8 of its time probing any computers on the Internet at random (global, fully-mixed spreading) and the rest of the time probing computers located in local area networks (local, fully-mixed spreading) \cite{Moore_2002}. 
Today information is propagated in society
via mass media (TV, newspaper, posters) as well as online social media (Facebook, Twitter and emails).
Mass media (global, fully-mixed spreading) can potentially deliver
the information to a big audience, but the effectiveness
of information transmission at an individual level may be small
(for example, its ability to alter the target individuals behaviour).
In contrast, social media (local, network spreading)
may have little or no access to the majority of people who are
not connected to the local group, but they provide rapid penetration of a selected target group with higher effectiveness. 

It is clear that hybrid epidemics are much more complex than simple epidemics. Their behaviour is affected not only by multiple spreading mechanisms that they use, but also by the population's overlaid structure on which they spread. Studying hybrid epidemics may provide crucial clues for better understanding of many real epidemics.

%Wang P, Gonz�lez MC, Hidalgo CA, Barab�si A-L (2009) Understanding the Spreading Patterns of Mobile Phone Viruses. Science 324(5930):1071 ?1076.

\subsection*{Previous works on hybrid epidemics}

Hybrid epidemics were initially studied as two levels of mixing in a population where nodes are mixed at both local and global levels \cite{Ball_1997}. Recently hybrid epidemics were studied as two levels of mixing in a network \cite{Kiss_Green_Kao_2006, Ball_Neal_2008, Estrada_2011}, in structured populations \cite{Vazquez_2007}, in structured households \cite{House_Keeling_2008, Ball_2012, Ma_2013},  and in a meta-population which consists of a number of weakly connected sub-populations\cite{Watts_2005,Colizza_Vespignani_2007,Mata_2013,Min_2013, Keeling_2010, Apolloni_2014}. Studies of epidemics in clustered networks \cite{Miller_2009,Tildesley_2010,Volz_2011} are also relevant to the hybrid epidemics. 

These previous works focused on analysing how a network's structure affects hybrid spreading. And most of them studied the {\em non-critically} hybrid epidemics, where at least one of the two spreading mechanisms alone can cause an infection outbreak and therefore the mix of two mechanisms is not a necessary condition for an epidemic outbreak. In this case, a hybrid epidemic using two spreading mechanisms is often less contagious than an epidemic using only one of the  mechanisms. \cite{Kiss_Green_Kao_2006, Wang_Jin_2013}. 

%\subsection*{critically hybrid epidemics} 

%critically hybrid epidemics are widely observed in nature and society. For example the HIV viruses spread between T-cells in human body via two routes: (1) the {\em cell-free} spreading, where susceptible cells are infected by the viruses (released by other infected cells) in body fluid; 
%(2) the {\em cell-to-cell} spreading, where infected cells infect (and transfer HIV viruses to) susceptible cells directly.
%Theoretical analysis suggests that if any of the two routes is blocked, the virus might be eliminated by human's immune system . However the mix of the two spreading routes makes HIV extremely contagious and difficult to cure. 
%
%Other examples include computer worms Red Cod II and Conficker which will be studied in detail in this paper. 

\subsection*{Our recent study on critically hybrid epidemics}

\begin{figure}[h]
\small\centering
\includegraphics[width=0.8\textwidth]{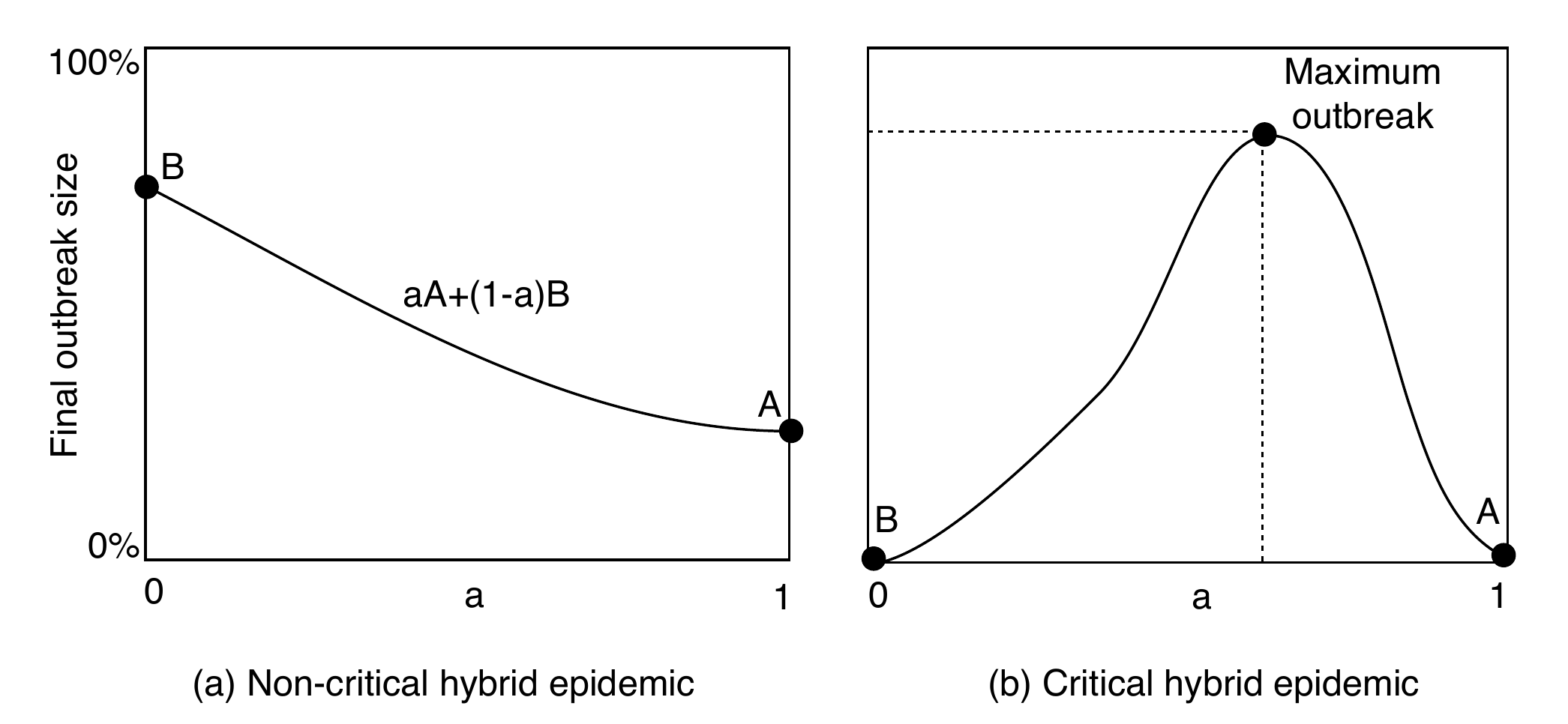}
\caption{\label{fig-Hybrid}Hybrid epidemics, where two spreading mechanisms A and B are mixed at the ratio of $\alpha$ to $(1-\alpha)$, where $0\leqslant\alpha \leqslant1$. (a)~Non-critically hybrid epidemic, where at least one of the two mechanisms can cause an outbreak by its own (i.e. when $\alpha=1$ or $\alpha=0$). (b)~critically hybrid epidemics, where each mechanism alone cannot cause any significant infection whereas a mix of them produces an epidemic outbreak. There exists an optimal $\alpha$ that produces the maximum outbreak.}
\end{figure}

We are interested in the {\em critically} hybrid epidemics, where each of the spreading mechanisms alone is not able to cause any significant infection whereas a combination of the mechanisms can cause an epidemic outbreak.
In this case, the mix of different spreading mechanisms is a critically condition for an outbreak 
(see \reffig{fig-Hybrid}).

Recently we proposed a generic model to study the critically hybrid epidemics \cite{Zhang_hm_2014}.
We considered an epidemic which spreads in a {\em meta-population} (consisting of many weakly connected {\em sub-populations}) using a mix of the following two typical spreading mechanisms.
(1) Fully-mixed spreading on the global level, i.e.\,infection between any two nodes in the meta-population.
(2) Network (or fully-mixed) spreading on the local level, i.e.\,infection between nodes within a sub-population where the internal topology of a sub-population is a network (or a fully-connected mesh). 
Each spreading mechanism has its own infection rate and an infected node recovers at a recovery rate. 
We define a parameter called the hybrid trade-off, $\alpha$, as the proportion of time that the epidemic devotes to the first spreading mechanism (or the probability of using the first spreading mechanism in a time unit). Thus the proportion of time spent on the second mechanism is $(1-\alpha)$. 

Our mathematical analysis and numerical simulations based on the model highlight the following two results. 
{Firstly, it is possible to mix two ineffective spreading mechanisms to produce a highly contagious epidemic, because the mix of the mechanisms can help to overcome the weakness of each mechanisms. 
Secondly, the threshold and the size of outbreak is critically determined by the hybrid trade-off $\alpha$. We also provided an analytical prediction of the optimal trade-off for the maximum outbreak size. 
}

\section*{Computer Worm Conficker}

In this paper we will analyse a critically hybrid epidemic, the computer worm Conficker, based on real measurement data. 
It is one of the most contagious computer worms on record. It erupted on the Internet on 21 November 2008 and infected millions of computers in just a few days \cite{Shin_2012}. 
The worm's ability to spread to such a large number of computers in so short a time and the fact \cite{ESET_Conficker} that it is still active on the Internet has caused serious concern. 

\begin{figure}[h]
\centering\small
\includegraphics[width=0.8\textwidth]{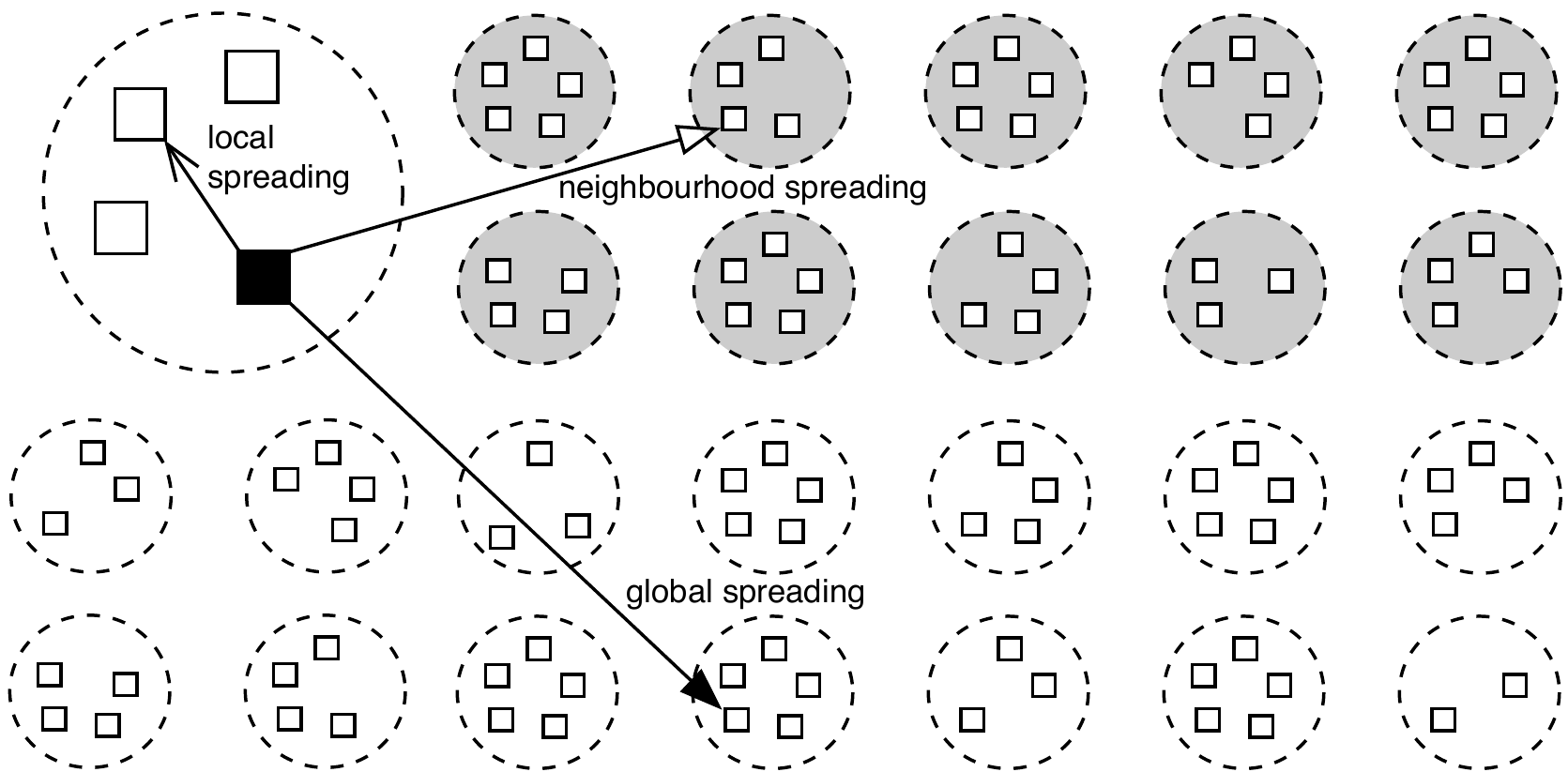} 
\caption{\label{fig-Conficker}Conficker's three probing strategies: (1) global spreading, where it probes any computer on the Internet at random;  (2) local spreading, where it probes computers in the same local network; (3) neighbourhood spreading, where it probes computers in ten neighbouring local networks.}
\end{figure}

Each computer on the Internet is associated with an Internet Protocol (IP) address. Conficker views the Internet as a meta-population, where computers are located in sub-populations, i.e. Local Area Network (LAN) consisting of computers whose IP addresses share the same prefix. 
According to the computer security company Symantec \cite{Eric_2010}, Conficker uses three spreading mechanisms (see \reffig{fig-Conficker}):
\begin{itemize}
\item {\em Global} spreading, where the worm probes computers with random IP addresses on the Internet; 
\item {\em Local} spreading, where the worm on an infected computer probes computers in the same Local Area Network (LAN) with the same IP address prefix; 
\item {\em Neighbourhood} spreading, where it probes computers in ten neighbouring LANs (with smaller consecutive IP address prefixes). 
\end{itemize}

Previous research on Conficker has studied the geographical distribution of infected IP addresses, the distribution of probing packet size~\cite{Shin_2012, Irwin_2012, Chiba_2013},  
and properties of the worm's global probing~\cite{Li_2009, Yao_2011}. 
The parameters of Conficker's hybrid spreading and how they affect the
epidemic dynamics of the worm can help explain why the worm is so contagious. But they have been hitherto little studied. 

\section*{Our Model of Conficker}

Here we use the notion of node to represent an IP address at which a computer (or computers) connects to the Internet.
For convenience, we call a sub-population (or a LAN) a subnet. 
A node must be in one of three states: {\em susceptible}, {\em infected}, and {\em recovered}~\cite{Newman_Book_2010}.
Initially only a small number of nodes are infected and all others are susceptible.
At each time step, an infected node attempts to spread the worm to susceptible nodes using one of the three probing strategies:
\begin{itemize} 
\item Global spreading with probability $\alpha_g$, where the worm probes nodes on the Internet at random with the global infection rate $\beta_g  \in [0,1]$.
\item Local spreading with probability $\alpha_l$, where it  probes nodes in the local subnet with the local infection rate $\beta_l \in [0,1]$; 
\item Neighbourhood spreading with the probability $\alpha_n$, where it probes nodes in ten neighbouring subnets with the neighbourhood infection rate $\beta_n \in [0,1]$; 
\end{itemize}
The mixing probabilities satisfy
$\alpha_g + \alpha_l + \alpha_n = 1.$

An infected node is recovered with recovery rate $\gamma\in[0,1]$. 
A recovered node remains recovered and cannot be infected again. 
Note that for mathematical analysis, the mixing probabilities could be incorporated into the infection rates. But we have treated them as separate parameters, considering that an infection rate reflects inherent properties of a computer worm in the context of a specific target population, whereas mixing probabilities are settings that can be easily modified in the worm's code. 
This is also the reason we use the mixing probabilities as controlling parameters in our study below and keep other parameters the same.

Only nodes that can potentially be infected by Conficker are relevant to our study. We call them the {\em relevant} nodes.
A subnet is relevant if it contains at least one relevant node. 
Irrelevant nodes include unused IP addresses and those computers that do not have the vulnerabilities that the worm can exploit.  Note that although the irrelevant nodes and subnets do not participate in the spreading of Conficker, they will be probed by the worm as the worm does not have the priori knowledge about which nodes are vulnerable. 

Let $n$ represent the total number of relevant nodes and $N$ the number of relevant subnets. The average number of relevant nodes in a subnet is $n_N=n/N$.
Let $N^+$ represent the average number of relevant subnets in ten neighbouring subnets.

At time $t$, the total number of susceptible, infected, and recovered nodes  are $S(t)$, $I(t)$, and $R(t)$, respectively.
Then the average number of infected nodes in a subnet is $I_N(t)=I(t)/N$, 
and the average number of infected nodes in ten neighbouring subnets is $I^+(t)=I_N(t)N^+$.
Hence on average a susceptible node can be infected via 
(1) {global probing} by $I(t)$ infected nodes in the Internet; 
(2) {local probing by $I_N(t)$ infected nodes in the local subnet; 
(3) {neighbourhood probing} by $I^+(t)$ infected nodes in the neighbouring subnets. 

The average probabilities that a susceptible node is \textit{not} infected by the global, local and neighbourhood probing, respectively, are: 
\begin{eqnarray}
P_g(t)&=& (1-\alpha_g\beta_g)^{I(t)}\nonumber\\ 
P_l(t)&=& (1-\alpha_l\beta_l)^{I_N(t)} \nonumber\\
P_n(t)&=& (1-\alpha_n\beta_n)^{I^+(t)}.
\label{eq-P1P2P3}
\end{eqnarray}
The average probability of \textit{not} being infected by any probing  is $P(t)=P_g(t)P_l(t)P_n(t)$. Thus the discrete evolution of Conficker spreading can be described as: 
\begin{eqnarray}
S(t+1)&=& S(t)-S(t)[1-P(t)] \nonumber \\
I(t+1)&=& I(t)+S(t)[1-P(t)]-\gamma I(t) \nonumber \\
R(t+1)&=& R(t)+\gamma I(t)
\label{eq-sirt+1}
\end{eqnarray}
where $S(t)[1-P(t)]$ is the number of new infections at  $t$.

\section*{Inferring Conficker Parameters From Data}

We infer the parameter values of our Conficker model from the Internet measurement data \cite{CAIDA_2008_aft, CAIDA_2008_bfr} collected by the Center for Applied Internet Data Analysis (CAIDA) in 2008. This is the only publicly available dataset that has captured the \textit{initial} outbreak process of the worm. The CAIDA Network Telescope project~\cite{CAIDA_2008_aft, CAIDA_2008_bfr} monitors Internet traffic sent to a large set of {\it unusable} IP addresses, which account for around $1/256$ of all addresses.  No legitimate traffic should be sent to these monitored addresses because they are not allocated for normal usage~\cite{Emile_Aben_2009}. 
Thus the traffic data captured by this project provides a good view on various abnormal behaviours on the Internet.  

When Conficker spreads on the Internet, its global spreading mechanism sends out probing packets to randomly generated IP addresses, some of which are unused IP addresses and therefore are monitored by the Network Telescope project. 
Conficke's probing packets are characterised by the Transmission Control Protocol  (TCP) with destination port number 445. This feature can be used to distinguish Conficker packets from other packets in the Network Telescope data. 

For each record of Conficker's probing packet, we are interested in two things: (1) the time when the packet is monitored by the Network Telescope project, and (2) the packet's source IP address, which gives the location of a Conficker-infected node. 
We ignore the destination address, as it is a randomly-generated, unused IP address.

We study the Network Telescope project's daily dataset collected on November 21, 2008, the day when Conficker broke out on the Internet. 
We use two earlier datasets collected on November 12 and 19, 2008 to filter out background `noise' that has been happening before the outbreak. 
That is, in the outbreak dataset, we discard  packets that were sent from any source address that had already sent packets to any of the unusable addresses in the two earlier datasets.
We use the prefix of /24 (i.e. IP address mask of 255.255.255.0) to distinguish different subnets~\cite{Shin_2012}. Our analysis uses a 10-minute window.

%%%%%%%%%%%%%%%%%%%%%%%%%%%

\subsection*{Step One: Inferring node status at a given time} 

We first infer the status of each node at time $t$ from the CAIDA data.
On the day of Conficker outbreak, all relevant nodes were initially susceptible.
In the analysis, we assume a node is just infected by the worm when we observe the first Conficker probing packet coming from it; and the node is recovered when we observe its last probing packet before the end of the day. 
\reffig{fig-SIRt} shows the number of susceptible, infected and recovered nodes as observed in a 10-minute window.

\begin{figure}[h]\centering\small
\includegraphics[width=0.6\textwidth]{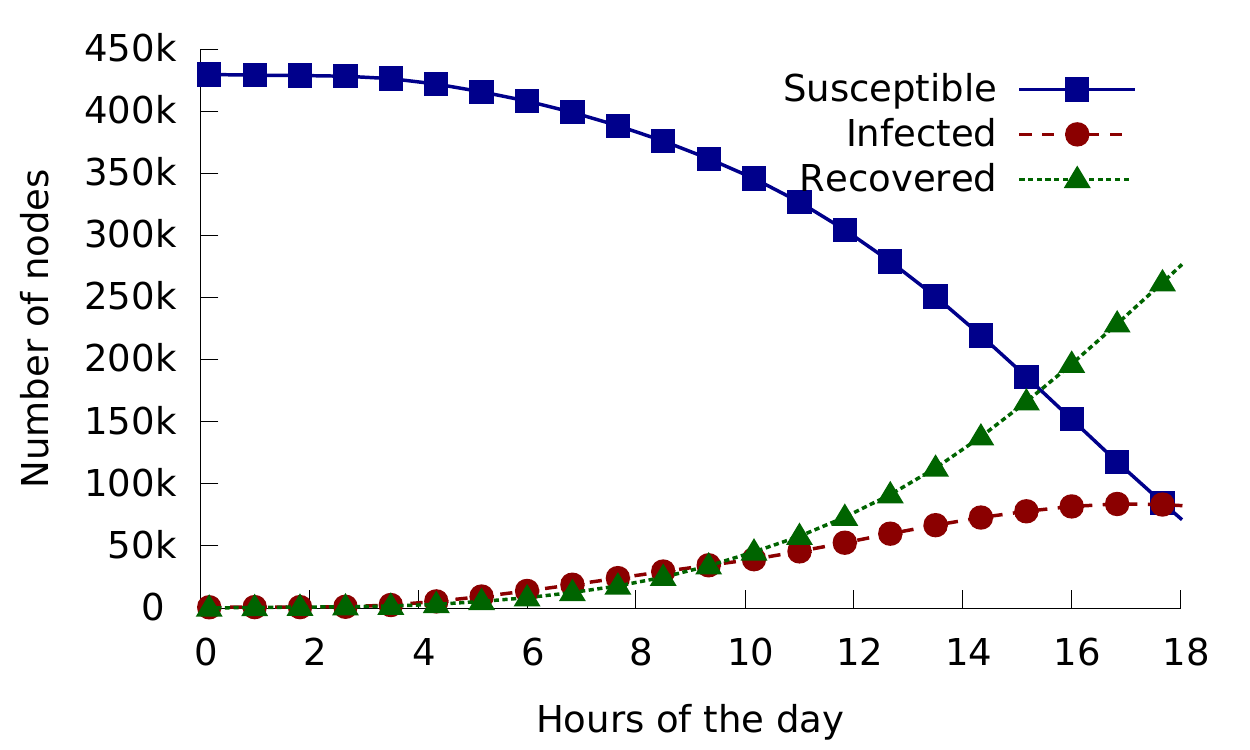}
\caption{\label{fig-SIRt}Numbers of susceptible nodes~$S(t)$, infected nodes~$I(t)$ and recovered nodes~$R(t)$ as a function of time $t$, as inferred from CAIDA's dataset on 21/Nov/2008, the day of Conficker's outbreak.}
\end{figure}

\subsection*{Step Two: Inferring new infections caused by each spreading mechanism} 

Let $dI_l(t)$, $dI_n(t)$ and $dI_g(t)$ represent the numbers of nodes that are newly infected through local, neighbourhood and global spreading, respectively, at time step $t$. 
%
%We will show in \refsec{sec-data-ana} that the following can be measured from the CAIDA data: $S(t)$, $I(t)$, $R(t)$, $dI_l(t)$, $dI_n(t)$, $dI_g(t)$, $N$, $n$, $n_N$ and $N^+$. Here we introduce our method for inferring Conficker's other parameters.  
%
%The infectious period of the node is from its infection to the recovery. 
%
Our analysis on the data shows that 84\% of new infections occurred within already infected subnets or their neighbourhood subnets, i.e. only 16\% of new infections appeared outside the reach of local and neighbourhood probing. This agrees with our understanding that local and neighbourhood probing are significantly more effective than global probing \cite{Shin_2012}. And 73\% of those new infections within the reach of local and neighbourhood probing (i.e. $73\%\times84\%$ of all new infections) occurred in already infected subnets. This indicates the local probing is more effective than neighbourhood probing. Based on the above analysis we can then approximately identify the probing mechanism that is responsible for a newly infected node by analysing the states of other relevant nodes at the time when the new infection happens.
\begin{itemize}
\item IF there is an infected node already in the same subnet, the new infection is caused by that infected node via local spreading.
\item ELSE IF there is an infected node in the ten neighbouring subnets, then the new infection is via neighbourhood spreading.
\item OTHERWISE, the newly infected node is infected via global spreading.
\end{itemize}
\reffig{fig-newly-infected} shows the inferred results, plotting the number of new infections caused by each spreading mechanism as a function of time.

\begin{figure}[h]
\small\centering
\includegraphics[width=0.6\textwidth]{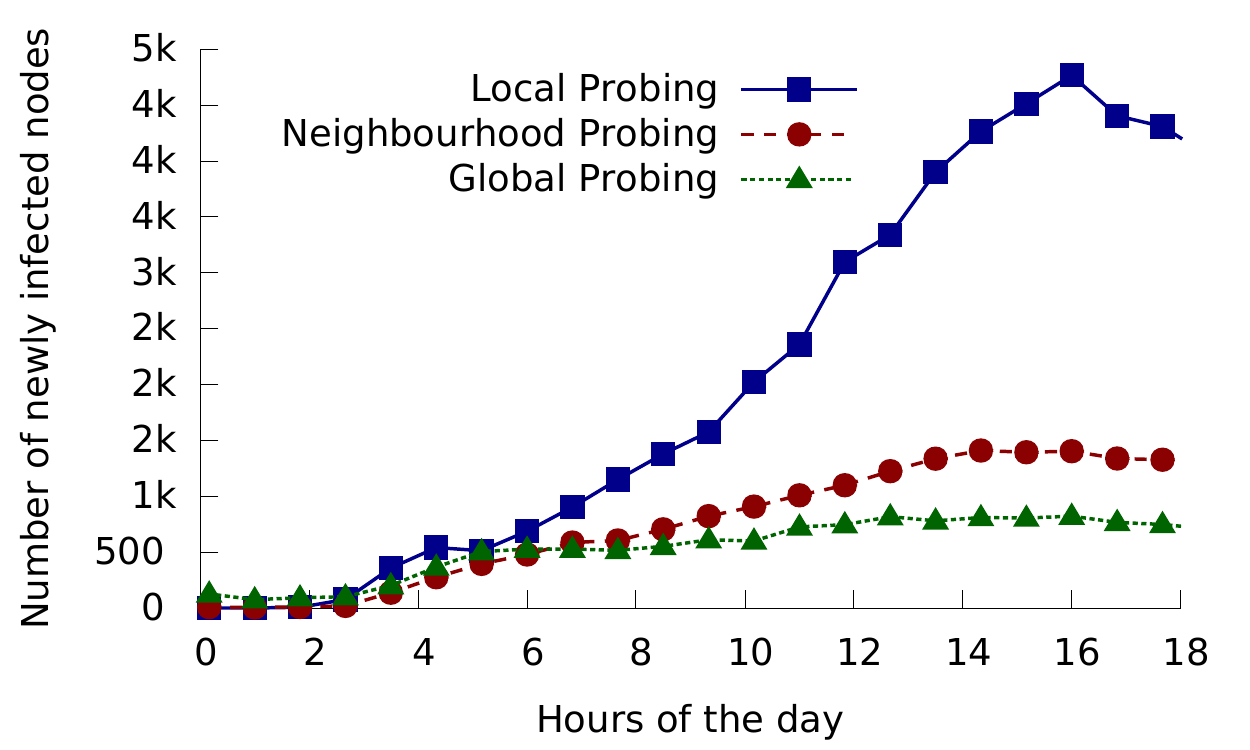}
\caption{\label{fig-newly-infected}Numbers of nodes newly infected by Conficker via each of the three spreading mechanisms in  10-minute windows on the day of Conficker's outbreak, as inferred from CAIDA's dataset on 21/Nov/2008.}
\end{figure}

%
%%%%%%%  !!!
%Note that for simplicity, in our model an infected node uses neighbourhood probing to scan its own neighbouring subnets (instead of those of a previous victim \cite{Eric_2010}).

\subsection*{Step Three: Inferring parameters of the Conficker model}\label{sec-method}

From the above result, we can calculate $I_N(t)$ and $I^+(t)$ according to their definitions. 
Then we calculate the average probabilities that a susceptible node at time $t$ is \textit{not} infected by the local, neighbourhood and global probing, respectively,  as: 
\begin{equation}\label{eq-PlPnPg-dI}
P_l(t)=1-\frac{dI_l(t)}{S(t)},P_n(t)=1-\frac{dI_n(t)}{S(t)},P_g(t)=1-\frac{dI_g(t)}{S(t)}.
\end{equation}
We define the effective infection rate as $b_l=\alpha_l \beta_l$, $b_n=\alpha_n \beta_n$ and $b_g = \alpha_g \beta_g$. 
Then we can use \refeq{eq-P1P2P3} and \refeq{eq-PlPnPg-dI} to calculate $b_l$ $b_n$ and $b_g$.

Let $\lambda$ denote the average total number of probes an infected node conducts during each time step. Then the average number of local, neighbourhood, and global probes in a time step are respectively $\alpha_l\lambda$, $\alpha_n\lambda$, and $\alpha_g\lambda$. 
The number of nodes (relevant and irrelevant) probed by the local, neighbourhood and global probing are 256, $10\times256$ and $2^{30}$ (it is not $2^{32}$ due to a bug in the worm's random number generation algorithm \cite{Eric_2010}). We can express the effective infection rates as:
\begin{equation}
\label{eq-lambda}
b_l=\alpha_l\lambda/256, ~b_n=\alpha_n\lambda/2560, ~b_g=\alpha_g\lambda/2^{30}.
\end{equation}
By solving \refeq{eq-lambda} together with $\alpha_g + \alpha_l + \alpha_n = 1$, we can obtain $\lambda$, $\alpha_l$, $\alpha_n$ and $\alpha_g$. 
Then we can obtain the infection rates of three spreading mechanisms as $\beta_l=b_l/\alpha_l$, $\beta_n=b_n/\alpha_n$, and $\beta_g=b_g/\alpha_g$. 
The recovery rate can be calculated as $\gamma=dR(t)/I(t)$, where $dR(t)=R(t+1) - R(t)$. 

%%%%%%%%%%%%%%%%%%%%%%%%%%%%
\subsection*{Inference results and evaluation}

The inferred values of the Conficker model parameters are shown in \reftab{tab-pars}, including the mixing probability $\alpha$ and the infection rate $\beta$ for three spreading mechanisms, the recover rate $\gamma$, the recovery time $\tau=1/\gamma$ which is the average time it takes for an infected node to recover, and the probing frequency $\lambda$. 
The parameter values are averaged over time windows between 4:00 and 16:00 when the spreading behaviour was stable. 
Computers are online and offline on a daily basis following a diurnal pattern \cite{David_2006}. We find that this factor only has a marginal impact on our results.

\begin{table}[h]\small\centering
\caption{Conficker parameters inferred from data$^1$} 
\label{tab-pars}
\begin{tabular}{rll}
\hline
Global spreading & $\alpha_g = 89.1\%$ & $\beta_g = 7.7\times10^{-8}$ \\
Local spreading&  $\alpha_l = 5.3\%$ & $\beta_l = 0.32$  \\
Neighbourhood spreading &  $\alpha_n = 5.6\%$ & $\beta_n = 0.032$ \\
Recovery rate& $\gamma=0.064$ &  \\
Recovery time & $\tau=156$ mins \\
Probing frequency& $\lambda=82.5$& \\
\hline
\end{tabular}\\
$^1$ All parameters are measured in a 10-minute window.
\end{table}

We observe in the data that the worm had infected in total $n$=430,135 nodes, which were located in $N$=92,267 subnets. On average, each subnet has $n_N$=4.7 relevant nodes, and $N^+$=4.3 of ten neighbouring subnets are relevant.

With these parameter values, we can use our Conficker model (see \refeq{eq-sirt+1}) to theoretically predict the worm's outbreak process. 
As measured from the data, the number of nodes in the three statuses were $S=423,899$, $I=3,945$, and $R=2,291$ at 4:00am. Our prediction starts from 4.00am and uses these numbers as the initial condition. 
As shown in \reffig{fig-fit}, our model's predictions closely match the measurement data.

\begin{figure}[h]
\centering\small
\includegraphics[width=0.6\textwidth]{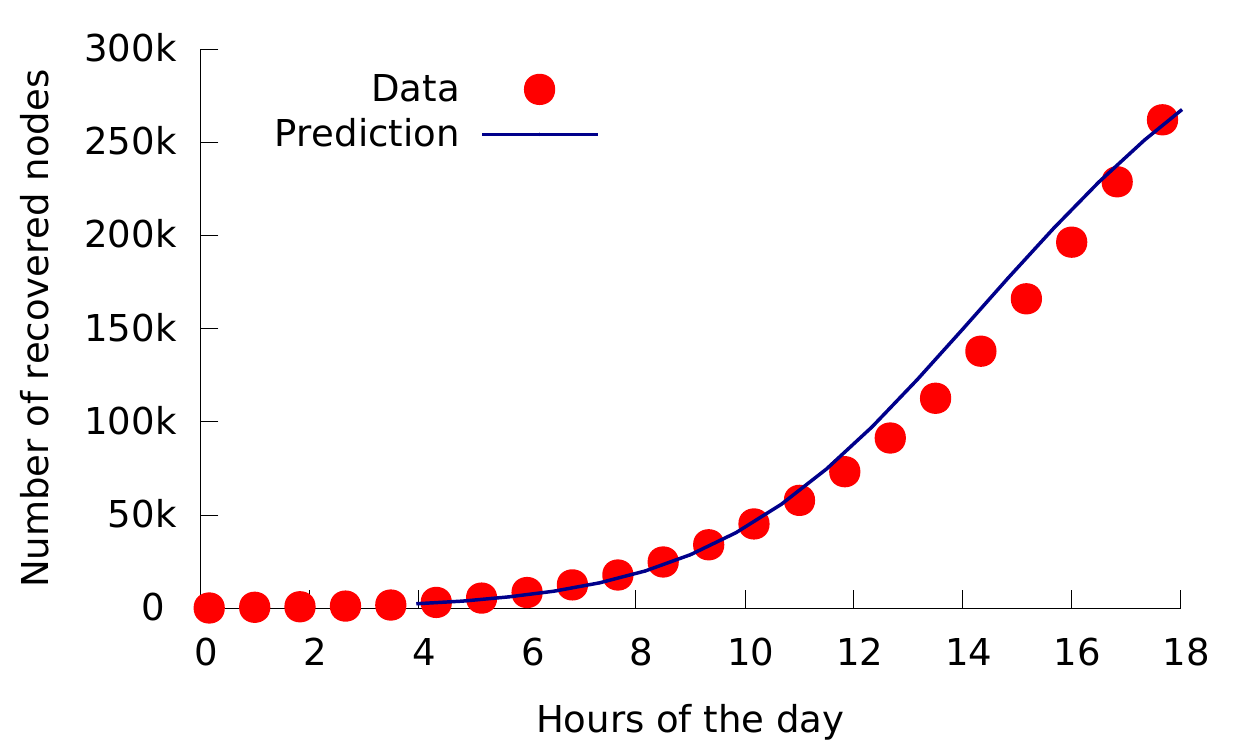}
\caption{\label{fig-fit}The outbreak of computer worm Conficker. Points are measured from Network Telescope's dataset collected on the outbreak day. Curve is theoretical prediction from our Conficker model using the inferred parameters.}
\end{figure}

The inferred parameters are in agreement with our expectations. For example the local spreading has a high infection rate because if a computer is already infected, then other computers in the same subnet are likely to have a similar computer system and thus are also likely to be vulnerable to the worm. 
By comparison, global spreading has an extremely low infection rate. On average, 
more than 10 million global probings will produce only a single new infection. 
On average an infected node retains its status for 2.5 hours (156 mins) before it recovers (e.g.~switched off or updated with new anti-virus database). The worm only sends out 8 probing packets per minute. Such a deliberately low probing rate helps the worm to evade a computer's or network's security systems.

\section*{Analysis on Conficker's Hybrid Spreading}

%%%%%%%%%%%%%%%%%%
\subsection*{Mix of TWO spreading mechanisms}

We run simulations using our Conficker model with the parameter values inferred above. The simulation network has 100k subnets. Each subnet contains 5 relevant nodes and has 4 relevant adjacent subnets. This topology setting resembles Conficker's spreading network observed in the data. Initially two random nodes are infected. The only controlling parameter is the mixing probabilities of the spreading mechanisms. Simulation results on mix of two spreading mechanisms are shown in \reffig{fig:conficker-2}.

\begin{figure}[h]
	\centering\small
	\begin{overpic}[width=0.35\textwidth]{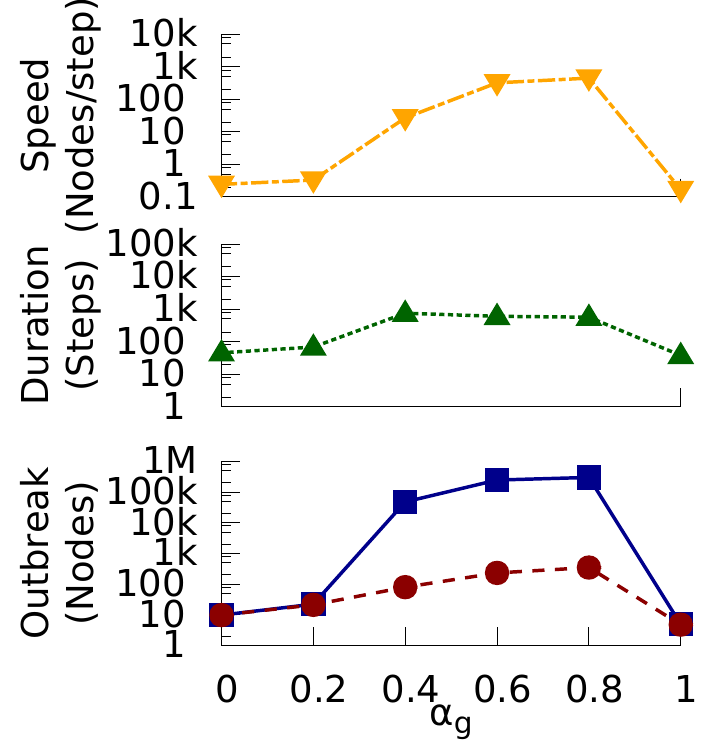}
		\put(27,96){(a) global and local}
	\end{overpic}	
	\begin{overpic}[width=0.25\textwidth]{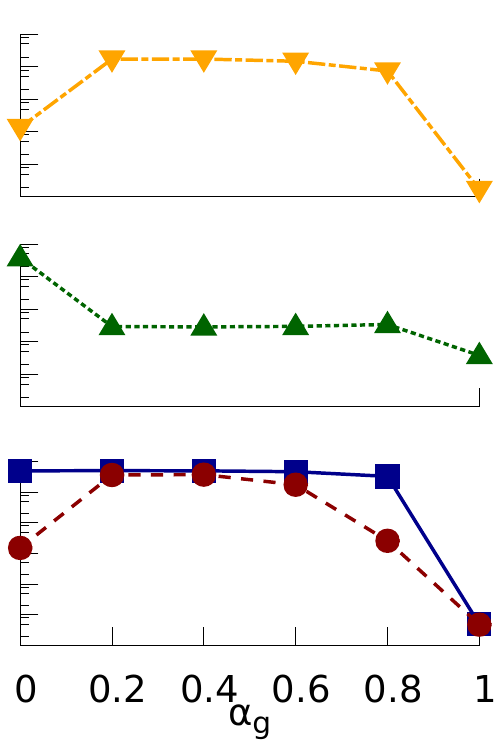}
		\put(0,96){(b) global and neighbourhood}
	\end{overpic}	
	\begin{overpic}[width=0.25\textwidth]{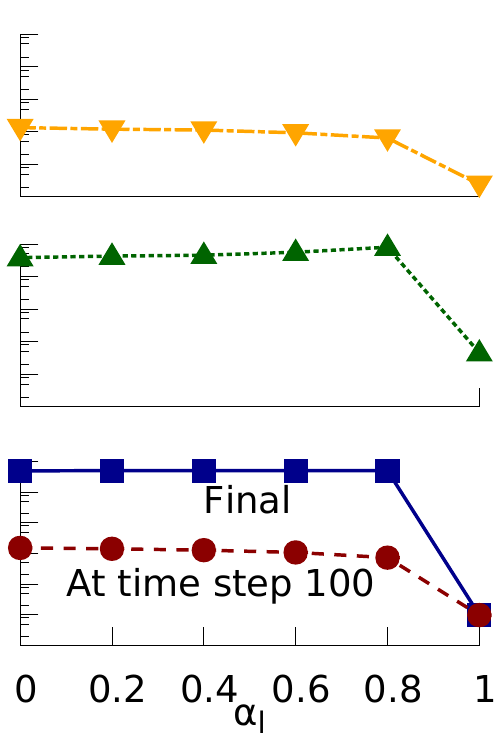}
		\put(0,96){(c) local and neighbourhood}
	\end{overpic}
\caption{\label{fig:conficker-2}Simulation results for the mix of Conficker's two spreading mechanisms with different mixing probabilities. 
		(a) Mix of global ($\alpha_g$) and local ($1-\alpha_g$) mechanisms;
		(b) Mix of global ($\alpha_g$) and neighbourhood (1-$\alpha_g$) mechanisms; 
		(c) Mix of local ($\alpha_l$) and neighbourhood (1-$\alpha_l$) mechanisms. In each case we measure the outbreak size, the total duration of the spreading, and the speed of spreading. The outbreak results include both the final outbreak size (square) and the outbreak size at time step 100 (filled circle). Each data point is averaged over 100 runs of a simulation. Note the y axes are all logarithmic.
	}
\end{figure}

\reffig{fig:conficker-2}a shows that as explained above, global spreading or local spreading alone cannot cause an outbreak, whereas a mixture at a ratio of 0.8 to 0.2 produces a large and rapid outbreak. 
\reffig{fig:conficker-2}b shows that the neighbourhood spreading alone ($\alpha_g=0$) can cause a large, but very slow outbreak, whereas the mix of neighbourhood spreading with just a small amount of global spreading can dramatically accelerate the spreading process. 
\reffig{fig:conficker-2}c shows that adding local spreading to neighbourhood spreading slows down the spreading process considerably. When they are mixed at the ratio of 0.8 to 0.2, the spreading reaches the same final outbreak size but the whole process lasts for the longest time.

%%%%%%%%%%%%%%%%%%
\subsection*{Mix of THREE spreading mechanisms}

Simulation results on mixing three spreading mechanisms are shown in \reffig{fig:conficker-3}. 
\reffig{fig:conficker-3}a shows it is not difficult to achieve a large final outbreak size when the three mechanisms are all present and neither local spreading nor global spreading is dominant.
\reffig{fig:conficker-3}b shows spreading will last for longer time if there is less global probing. 
\reffig{fig:conficker-3}c shows that the most contagious variation of the worm is a mix of global, local and neighbourhood spreading at the probabilities of 0.4, 0.2 and 0.4 (see circle on the plot), which causes the largest final outbreak with the highest spreading speed.

\begin{figure}[H]
\small\centering
\begin{overpic}[width=0.65\textwidth]{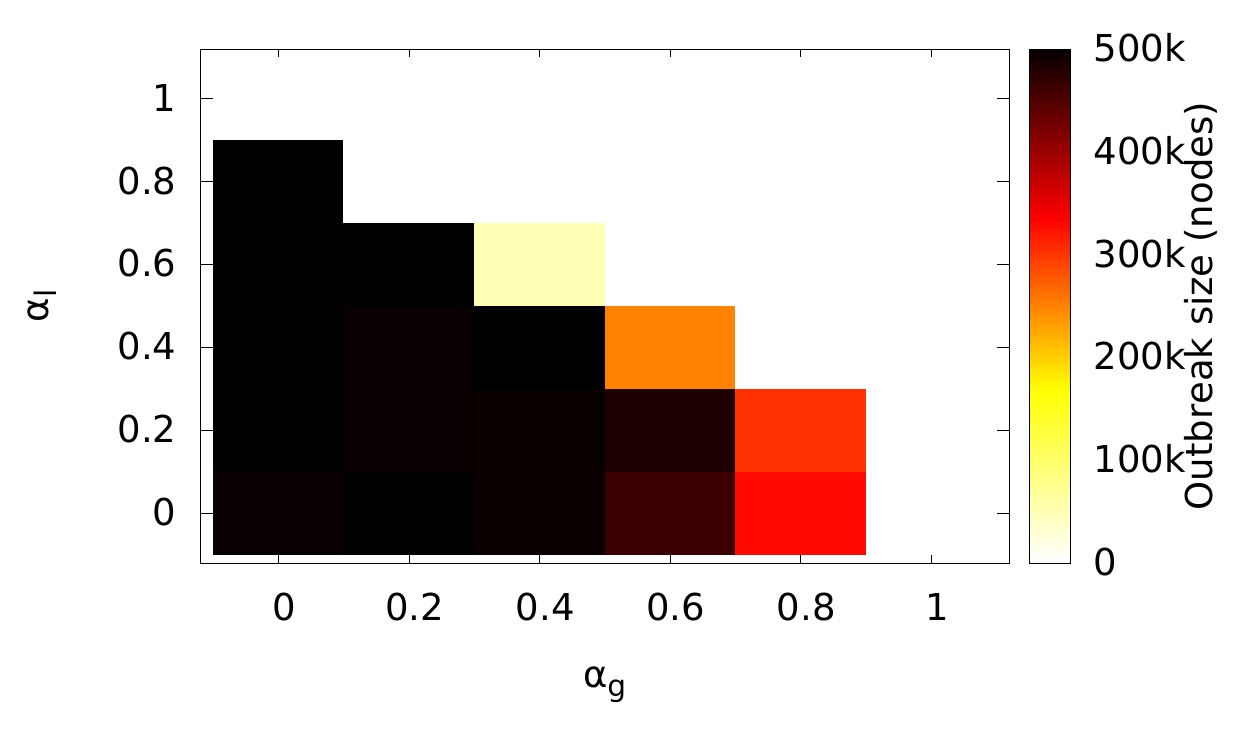}
	\put(38,50){(a) Outbreak size}
\end{overpic}
\begin{overpic}[width=0.65\textwidth]{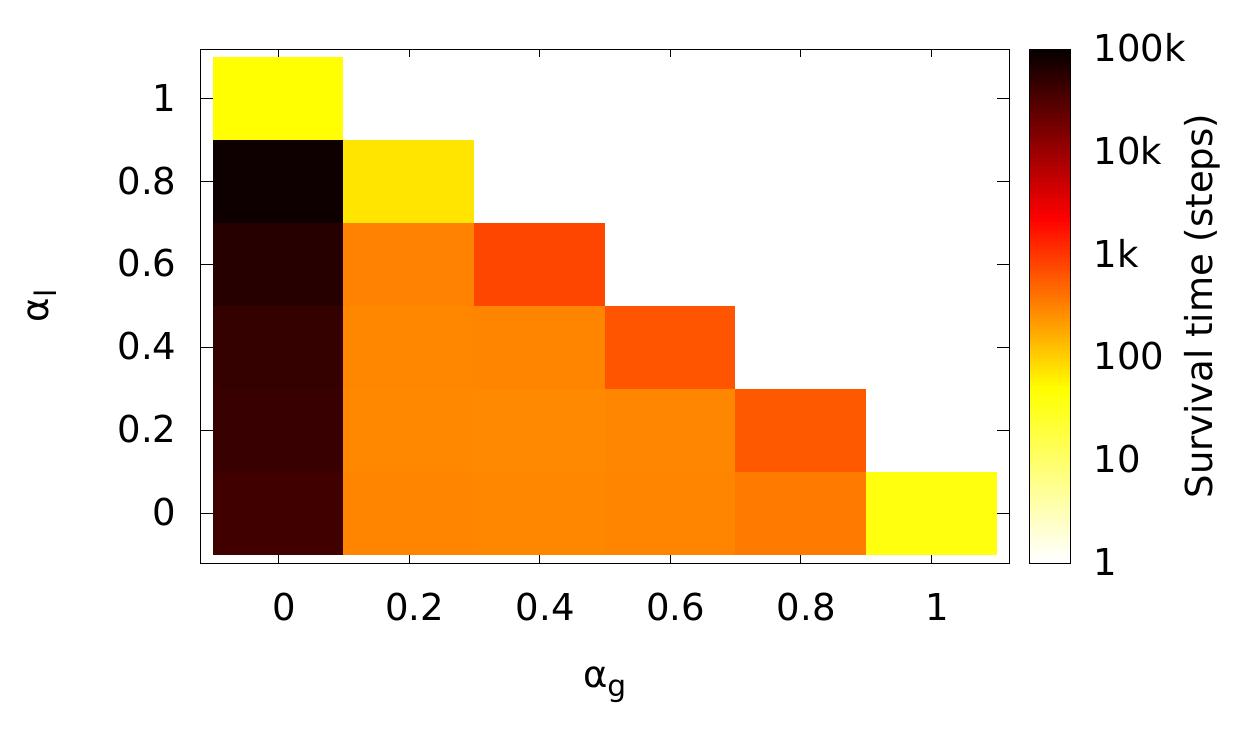}
	\put(38,50){(b) Survival time}
\end{overpic}
\begin{overpic}[width=0.65\textwidth]{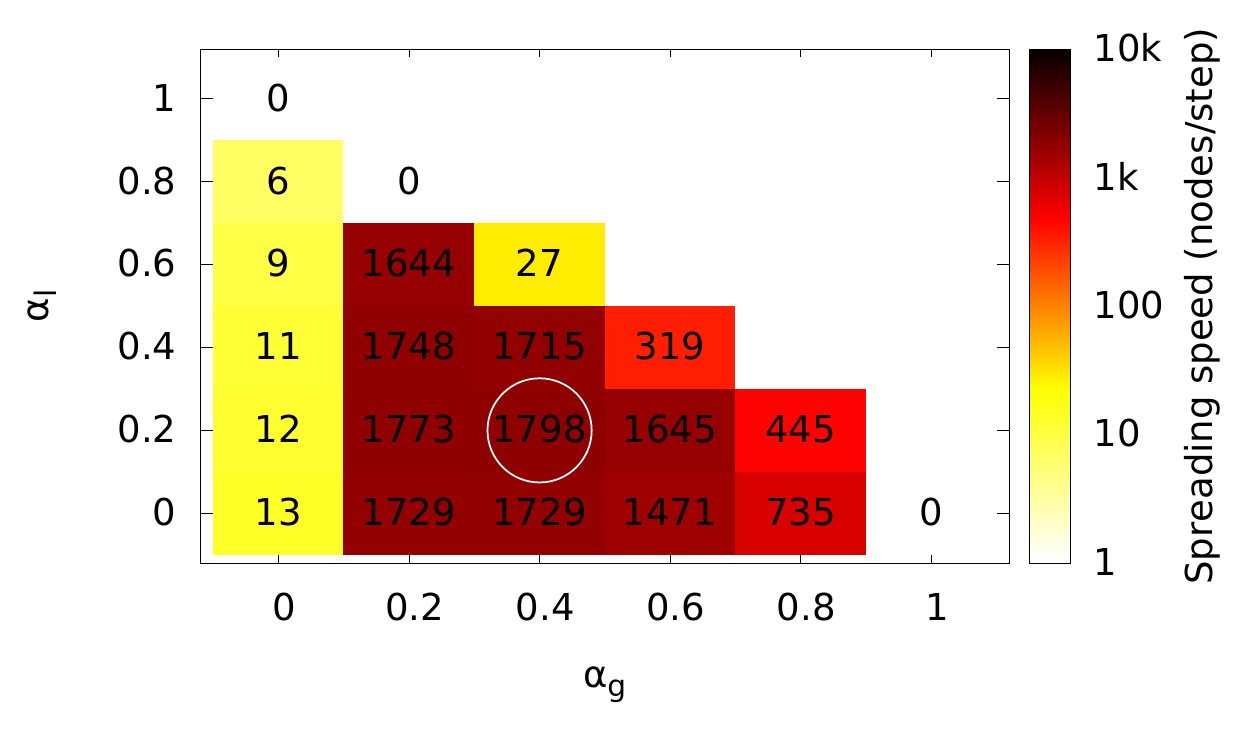}
	\put(38,50){(c) Spreading speed}
\end{overpic}
\caption{\label{fig:conficker-3}Simulation results when three of Conficker's spreading mechanisms are mixed at different probabilities. Spreading properties shown include the final outbreak size, the survival time and the spreading speed (see colour maps) as functions of the mixing probabilities of global spreading $\alpha_g$ (x axis) and local spreading $\alpha_l$ (y axis), where the mixing probability of neighbourhood spreading is $\alpha_n=1-\alpha_g-\alpha_l$.}
\end{figure}

%%%%%%%%%%%%%%%%%%%%%

\section*{Discussion}
In this study, we infer the epidemic spreading parameters of the Conficker worm from observed data collected during the first few hours of the epidemic. Simulations of worm spreading, based on these parameters, allow us to reach some important conclusions about the worm's use of hybrid spreading mechanisms.

\subsection*{Advantage of mixing hybrid spreading mechanism}

Conficker's global probing is extremely ineffective. 
The infection rate of global probing is many orders of magnitude smaller than the recovery rate. This means, if Conficker used only the global probing, it would not have caused any significant infection on the Internet at all. 

Local probing has a remarkably high infection rate, $\beta_l=0.32$, which means when an infected node conducts only local spreading, a susceptible node in the same subnet has an 1/3 chance of being infected in a step (10-mins).
However, local probing is confined within a subnet. If the worm used only the local probing, it would not have infected any other subnet apart from those initially containing infected nodes. 

Neighbourhood probing is constrained to a neighbourhood of ten subnets. 
It has a high infection rate because computers in adjacent IP address blocks often belong to the same organisation and they use similar computer systems and therefore have similar vulnerabilities that can be exploited by the worm. 
Since different nodes' neighbourhoods can partially overlap with each other, it is in theory possible for the worm to reach any node in the whole meta-population by using only the neighbourhood probing. Such process, however, would be extraordinarily slow as we have shown in \reffig{fig:conficker-2}b. 

In summary, if Conficker used only a single spreading mechanism, it would have vanished on the Internet without causing any significant impact. 

Thus the enormous outbreak of the worm lies in its ability to do two things. Firstly it needs to devote great efforts to explore every corner of the Internet to find a new vulnerable computer. Every new victim will open a new colony full of similar vulnerable computers. Secondly it needs to make the most out of each new colony. 

This is exactly what Conficker does. It allocates most of its time on global probing with a mixing probability of $\alpha=89\%$. This in a degree compensates the ineffectiveness of global probing. 
Although the worm allocates small amounts of time on local and neighbouring probing, their high infection rates allow them to exploit all possible victims in the subnets with efficiency. And all newly infected nodes will join the collective effort to flood the Internet with more global random probes. 

In short, the Conficker worm is an example of a critically hybrid epidemic. It can cause an enormous outbreak not because it has an advanced ability to exploit weaknesses of a computer, but because it has remarkable capability to discover all potentially vulnerable computers in the Internet, i.e. it is not the infectivity, but the hybrid spreading that makes Conficker one of the most infectious worms on record. 

%%%%%%%%%%%%%%
\subsection*{Implication of critically hybrid epidemics}

The analysis of critically hybrid epidemics such as Conficker has important general implications. Firstly, it demonstrates that it is possible to design a high impact epidemic based on mechanisms, each of relatively low efficiency. Indeed our result in \reffig{fig:conficker-3} suggests that Conficker could have had a larger outbreak with higher speed if it had used a different set of mixing probabilities, which  requires change of only a few lines of Conficker's program code. Hybrid mechanisms may therefore be ideal for rapid efficient penetration of a network, for example in the context of an advertising campaign or in order to promulgate important public health or security information. An interesting example might be the use of media campaigns (global spreading) where the reader or viewer is specifically requested to pass on a message via Twitter or Facebook to their ``local'' group contacts. 

Conversely, malicious hybrid epidemics can be extremely difficult to defend against, and many existing defence strategies may not be effective. For example immunising a selected portion of a local population in order to isolate and hence protect the vulnerable nodes will not be effective, because the vulnerable nodes can still be found by the worm through random global spreading. 

Another possible measure is to reduce the average time it takes for an infected node to recover, for example to speed up the release of anti-virus software updates or increase the frequency of security scanning on computers.
Our theoretical predictions (using \refeq{eq-sirt+1}) in \reffig{fig-tau} show that the final outbreak size (in terms of total recovered nodes) does not change significantly when the recovery time is reduced from 156 minutes to 140 or 120 minutes. In practice, even achieving such reductions  represents a remarkable technical challenge. It is clear from the discussion above that epidemics can spread with extremely low global infection rates (far below individual recovery rates), provided there is efficient local infection. The extremely efficient spreading achieved once a given subnet or set of subnets has been penetrated is therefore obviously a key determinant of the worm's outbreak \cite{Shin_2012}. Thus, defence strategies that focus on security co-operation between nodes with a local network neighbourhood (a ``neighbourhood watch'' strategy \cite{Shin_2012}) may be the key to future prevention of similar outbreaks.

\begin{figure}[h]
\centering\small
\includegraphics[width=0.65\textwidth]{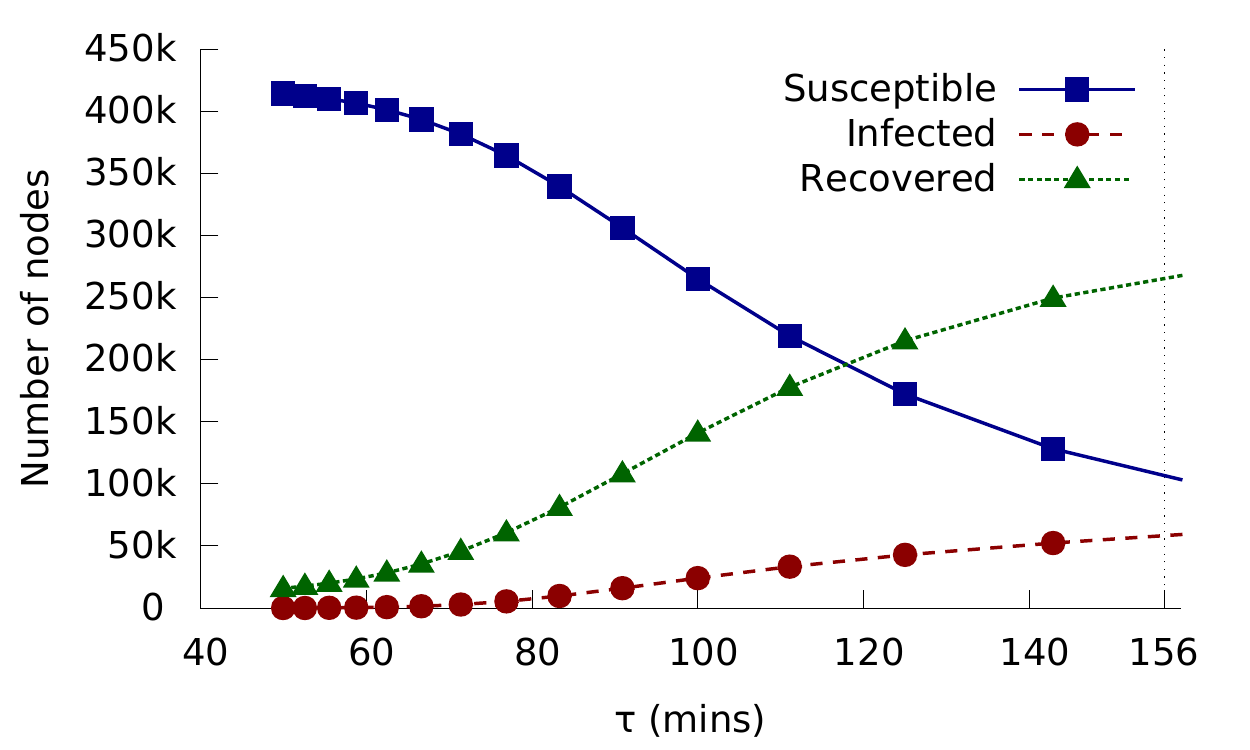}
\caption{\label{fig-tau}Predicted numbers of susceptible, infected and recovered nodes at 16:00 on the outbreak day as a function of the recovery time $\tau$, which is the average time for an infected node to recover. Conficker's recovery time is 156 minutes.}
\end{figure}

\subsection*{Our Conficker model}

The Conficker worm can be described as a discrete model or a continuous model. The two modelling approaches should give the same prediction results of the spreading dynamics of the worm. In this work we used a discrete approach to model the Conficker worm for three reasons. Firstly the model's parameters can be defined with clear physical meanings. Secondly we can directly calculate the parameters' values from the CAIDA measurement data. Lastly it is more convenient to run simulations with a discrete model. If a continuous model were used, the model parameters would be defined differently with less clear physical meanings, and their values would have to be obtained through iterative data fitting.

In our Conficker model, we set the local and global population as fully mixed, because this is how the Conficker worm perceives the structure of the Internet. We considered more complex network structures in a separate work \cite{Zhang_hm_2014} where we studied hybrid epidemics in general.

\section*{Conclusion}

Our study uses data collected during the first day of the Conficker epidemic to parametrise a hybrid model to capture the worm's spreading behaviour. The study highlights the importance of mixing different modes of spreading in order to achieve large, rapid and  sustained epidemics, and  suggests that the trade-off between the different modes of spreading will be critical in determining the epidemic outcome.

\section*{Acknowledgments}

%\section*{Author Contributions}
%C.Z., S.Z. and B.M.C. designed the experiments, analysed the results and wrote the paper; C.Z. designed the algorithms and performed the experiments.

\bibliographystyle{plos2015_pp.bst}
\bibliography{references}

\begin{thebibliography}{10}

\bibitem{House_2012}
House T.
\newblock Modelling epidemics on networks.
\newblock Contemp Phys. 2012;53(3):213.

\bibitem{Rock_2014}
Rock K, Brand S, Moir J, Keeling MJ.
\newblock Dynamics of infectious diseases.
\newblock Rep Prog Phys. 2014 Feb;77(2):026602.

\bibitem{R_Zheng_2014_3}
Zheng Y, Huang B, Wang Y, Sun X, Jin Z, Zhang J, et~al.
\newblock Transmission dynamics and control for a brucellosis model in Hinggan
  League of Inner Mongolia, China.
\newblock Math Biosci Eng. 2014 Jun;11(5):1115--1137.

\bibitem{R_Nie_2014}
Nie J, Sun GQ, Sun XD, Zhang J, Wang N, Wang YM, et~al.
\newblock Modeling the transmission dynamics of dairy cattle brucellosis in
  jilin province, china.
\newblock J Biol Syst. 2014 May;22(04):533--554.

\bibitem{R_Li_2014}
Li MT, Sun GQ, Wu YF, Zhang J, Jin Z.
\newblock Transmission dynamics of a multi-group brucellosis model with mixed
  cross infection in public farm.
\newblock Appl Math Comput. 2014 Jun;237:582--594.

\bibitem{Zou_2006}
Zou CC, Towsley D, Gong W.
\newblock On the performance of Internet worm scanning strategies.
\newblock Perform Eval. 2006 Jul;63(7):700--723.

\bibitem{Shin_2012}
Shin S, Gu G, Reddy N, Lee CP.
\newblock A Large-Scale Empirical Study of Conficker.
\newblock IEEE Trans Inf Forensics Secur. 2012 Apr;7(2):676--690.

\bibitem{Yu_2015}
Yu S, Gu G, Barnawi A, Guo S, Stojmenovic I.
\newblock Malware Propagation in Large-Scale Networks.
\newblock IEEE Trans Knowl Data Eng. 2015 Jan;27(1):170--179.

\bibitem{Ren_2014}
Ren Z, Liu W, Zhou X, Fang J, Chen Q.
\newblock Summary-Vector-Based Effective and Fast Immunization for
  Epidemic-Based Routing in Opportunistic Networks.
\newblock IEEE Commun Lett. 2014 Jul;18(7):1183--1186.

\bibitem{Chen_2014}
Chen PY, Cheng SM, Chen KC.
\newblock Optimal Control of Epidemic Information Dissemination Over Networks.
\newblock IEEE T Cybern. 2014 Dec;44(12):2316--2328.

\bibitem{Sahneh_2014}
Sahneh Fd, Chowdhury Fn, Brase G, Scoglio Cm.
\newblock Individual-based Information Dissemination in Multilayer Epidemic
  Modeling.
\newblock Math Model Nat Phenom. 2014 Jan;9(02):136--152.

\bibitem{R_Zhang_2014_2}
Zhang J, Sun GQ, Sun XD, Hou Q, Li M, Huang B, et~al.
\newblock Prediction and Control of Brucellosis Transmission of Dairy Cattle in
  Zhejiang Province, China.
\newblock PLoS ONE. 2014 Nov;9(11):e108592.

\bibitem{R_Zhang_2014}
Zhang J, Jin Z, Sun GQ, Sun XD, Wang YM, Huang B.
\newblock Determination of Original Infection Source of H7N9 Avian Influenza by
  Dynamical Model.
\newblock Sci Rep. 2014 May;4:4846.

\bibitem{Moore_2003}
Moore D, Paxson V, Savage S, Shannon C, Staniford S, Weaver N.
\newblock Inside the Slammer worm.
\newblock IEEE Secur Priv. 2003 Aug;1(4):33--39.

\bibitem{Shannon_Moore_2004}
Shannon C, Moore D.
\newblock The spread of the Witty worm.
\newblock IEEE Secur Priv. 2004 Aug;2(4):46--50.

\bibitem{Zhang_hm_2014}
Zhang C, Zhou S, Cox IJ, Chain BM. Optimizing Hybrid Spreading in
  Metapopulations; 2014.
\newblock Preprint. Available: arXiv:1409.7291. Accessed 10 Feb 2015.

\bibitem{Eric_2010}
Chien E. Downadup: Attempts at Smart Network Scanning; 2010.
\newblock Available:
  \url{http://www.symantec.com/connect/blogs/downadup-attempts-smart-network-scanning}.
\newblock Accessed Dec 2014.

\bibitem{CAIDA_2008_aft}
{Center for Applied Internet Data Analysis}. The CAIDA UCSD Network Telescope
  ``Three Days Of Conficker''; 2008.
\newblock Available:
  \url{http://www.caida.org/data/passive/telescope-3days-conficker_dataset.xml}.
\newblock Accessed Dec 2014.

\bibitem{CAIDA_2008_bfr}
{Center for Applied Internet Data Analysis}. The CAIDA UCSD Network Telescope
  ``Two Days in November 2008'' Dataset; 2008.
\newblock Available:
  \url{http://www.caida.org/data/passive/telescope-2days-2008_dataset.xml}.
\newblock Accessed Dec 2014.

\bibitem{Newman_Book_2010}
Newman M.
\newblock Networks: An Introduction.
\newblock Oxford University Press, USA; 2010.

\bibitem{Keeling_Eames_2005}
Keeling M, Eames K.
\newblock Networks and epidemic models.
\newblock J R Soc Interface. 2005 Sep;2(4):295--307.

\bibitem{Anderson_1991}
Anderson RM.
\newblock Discussion: The Kermack-McKendrick epidemic threshold theorem.
\newblock Bull Math Biol. 1991 Mar;53(1-2):1--32.

\bibitem{Pastor-Satorras_Vespignani_2001}
Pastor-Satorras R, Vespignani A.
\newblock Epidemic Spreading in Scale-Free Networks.
\newblock Phys Rev Lett. 2001 Apr;86(14):3200.

\bibitem{R_Wang_2013}
Wang L, Wang Z, Zhang Y, Li X.
\newblock How human location-specific contact patterns impact spatial
  transmission between populations?
\newblock Sci Rep. 2013 Mar;3:1468.

\bibitem{R_Wang_2013_2}
Wang L, Zhang Y, Wang Z, Li X.
\newblock The impact of human location-specific contact pattern on the sir
  epidemic transmission between populations.
\newblock Int J Bifurcation Chaos. 2013 May;23(05):1350095.

\bibitem{Balcan_2009b}
Balcan D, Colizza V, Gon\c{c}alves B, Hu H, Ramasco JJ, Vespignani A.
\newblock Multiscale mobility networks and the spatial spreading of infectious
  diseases.
\newblock Proc Natl Acad Sci USA. 2009 Dec;.

\bibitem{Wang_2009}
Wang P, Gonz\'{a}lez MC, Hidalgo CA, Barab\'{a}si AL.
\newblock Understanding the Spreading Patterns of Mobile Phone Viruses.
\newblock Science. 2009 May;324(5930):1071--1076.

\bibitem{Balcan_2011}
Balcan D, Vespignani A.
\newblock Phase transitions in contagion processes mediated by recurrent
  mobility patterns.
\newblock Nat Phys. 2011 Jul;7(7):581--586.

\bibitem{Meloni_2011}
Meloni S, Perra N, Arenas A, G\'{o}mez S, Moreno Y, Vespignani A.
\newblock Modeling human mobility responses to the large-scale spreading of
  infectious diseases.
\newblock Sci Rep. 2011 Aug;1.

\bibitem{R_Sun_2007}
Sun G, Jin Z, Liu QX, Li L.
\newblock Pattern formation in a spatial S-I model with non-linear incidence
  rates.
\newblock J Stat Mech. 2007 Nov;2007(11):P11011.

\bibitem{R_Sun_2008}
Sun GQ, Jin Z, Liu QX, Li L.
\newblock Chaos induced by breakup of waves in a spatial epidemic model with
  nonlinear incidence rate.
\newblock J Stat Mech. 2008 Aug;2008(08):P08011.

\bibitem{R_Sun_2010}
Sun GQ, Liu QX, Jin Z, Chakraborty A, Li BL.
\newblock Influence of infection rate and migration on extinction of disease in
  spatial epidemics.
\newblock J Theor Biol. 2010 May;264(1):95--103.

\bibitem{R_Sun_2012}
Sun GQ.
\newblock Pattern formation of an epidemic model with diffusion.
\newblock Nonlinear Dyn. 2012 Jan;69(3):1097--1104.

\bibitem{Moore_2002}
Moore D, Shannon C, Claffy KC.
\newblock Code-Red: a case study on the spread and victims of an internet worm.
\newblock In: Proceedings of the 2nd ACM SIGCOMM Workshop on Internet
  measurment. IMW. ACM; 2002. pp. 273--284.

\bibitem{Ball_1997}
Ball F, Mollison D, Scalia-Tomba G.
\newblock Epidemics with two levels of mixing.
\newblock Ann Appl Probab. 1997 Feb;7(1):46--89.

\bibitem{Kiss_Green_Kao_2006}
Kiss IZ, Green DM, Kao RR.
\newblock The effect of contact heterogeneity and multiple routes of
  transmission on final epidemic size.
\newblock Math Biosci. 2006 Sep;203(1):124.

\bibitem{Ball_Neal_2008}
Ball F, Neal P.
\newblock Network epidemic models with two levels of mixing.
\newblock Math Biosci. 2008 Mar;212(1):69.

\bibitem{Estrada_2011}
Estrada E, Kalala-Mutombo F, Valverde-Colmeiro A.
\newblock Epidemic spreading in networks with nonrandom long-range
  interactions.
\newblock Phys Rev E. 2011;84(3):036110.

\bibitem{Vazquez_2007}
Vazquez A.
\newblock Epidemic outbreaks on structured populations.
\newblock J Theor Biol. 2007 Mar;245(1):125--129.

\bibitem{House_Keeling_2008}
House T, Keeling MJ.
\newblock Deterministic epidemic models with explicit household structure.
\newblock Math Biosci. 2008 May;213(1):29--39.

\bibitem{Ball_2012}
Ball F.
\newblock An SIR epidemic model on a population with random network and
  household structure, and several types of individuals.
\newblock Adv Appl Probab. 2012 Mar;44(1):63--86.

\bibitem{Ma_2013}
Ma J, Driessche Pvd, Willeboordse FH.
\newblock Effective degree household network disease model.
\newblock J Math Biol. 2013 Jan;66(1-2):75--94.

\bibitem{Watts_2005}
Watts DJ, Muhamad R, Medina DC, Dodds PS.
\newblock Multiscale, resurgent epidemics in a hierarchical metapopulation
  model.
\newblock Proc Natl Acad Sci USA. 2005;102(32):11157--11162.

\bibitem{Colizza_Vespignani_2007}
Colizza V, Vespignani A.
\newblock Invasion Threshold in Heterogeneous Metapopulation Networks.
\newblock Phys Rev Lett. 2007 Oct;99(14):148701.

\bibitem{Mata_2013}
Mata AS, Ferreira SC, Pastor-Satorras R.
\newblock Effects of local population structure in a reaction-diffusion model
  of a contact process on metapopulation networks.
\newblock Phys Rev E. 2013 Oct;88(4):042820.

\bibitem{Min_2013}
Min Y, Jin X, Ge Y, Chang J.
\newblock The Role of Community Mixing Styles in Shaping Epidemic Behaviors in
  Weighted Networks.
\newblock PLoS ONE. 2013 Feb;8(2):e57100.

\bibitem{Keeling_2010}
Keeling MJ, Danon L, Vernon MC, House TA.
\newblock Individual identity and movement networks for disease
  metapopulations.
\newblock Proc Natl Acad Sci USA. 2010 May;107(19):8866--8870.

\bibitem{Apolloni_2014}
Apolloni A, Poletto C, Ramasco JJ, Jensen P, Colizza V.
\newblock Metapopulation epidemic models with heterogeneous mixing and travel
  behaviour.
\newblock Theor Biol Med Model. 2014 Jan;11(1):3.

\bibitem{Miller_2009}
Miller JC.
\newblock Spread of infectious disease through clustered populations.
\newblock J R Soc Interface. 2009 Dec;6(41):1121--1134.

\bibitem{Tildesley_2010}
Tildesley MJ, House TA, Bruhn MC, Curry RJ, O'Neil M, Allpress JLE, et~al.
\newblock Impact of spatial clustering on disease transmission and optimal
  control.
\newblock Proc Natl Acad Sci USA. 2010 Jan;107(3):1041--1046.

\bibitem{Volz_2011}
Volz EM, Miller JC, Galvani A, Ancel~Meyers L.
\newblock Effects of Heterogeneous and Clustered Contact Patterns on Infectious
  Disease Dynamics.
\newblock PLoS Comput Biol. 2011 Jun;7(6):e1002042.

\bibitem{Wang_Jin_2013}
Wang Y, Jin Z.
\newblock Global analysis of multiple routes of disease transmission on
  heterogeneous networks.
\newblock Physica A. 2013 Sep;392(18):3869--3880.

\bibitem{ESET_Conficker}
{ESET Virusradar}. Win32/Conficker Charts; 2014.
\newblock Available:
  \url{http://www.virusradar.com/en/Win32_Conficker/chart/week}.
\newblock Accessed Dec 2014.

\bibitem{Irwin_2012}
Irwin B.
\newblock A network telescope perspective of the Conficker outbreak.
\newblock In: Information Security for South Africa; 2012. pp. 1--8.

\bibitem{Chiba_2013}
Chiba D, Tobe K, Mori T, Goto S.
\newblock Analyzing Spatial Structure of IP Addresses for Detecting Malicious
  Websites.
\newblock Journal of Information Processing. 2013;21(3):539--550.

\bibitem{Li_2009}
Li R, Gan L, Jia Y.
\newblock Propagation Model for Botnet Based on Conficker Monitoring.
\newblock In: International Symposium on Information Science and Engineering;
  2009. pp. 185--190.

\bibitem{Yao_2011}
Yao Y, Xiang Wl, Guo H, Yu G, Gao FX.
\newblock Diurnal Forced Models for Worm Propagation Based on Conficker
  Dataset.
\newblock In: International Conference on Multimedia Information Networking and
  Security; 2011. pp. 431--435.

\bibitem{Emile_Aben_2009}
Aben E. Conficker/Conflicker/Downadup as seen from the UCSD Network Telescope;
  2009.
\newblock Available:
  \url{http://www.caida.org/research/security/ms08-067/conficker.xml}.
\newblock Accessed Dec 2014.

\bibitem{David_2006}
Dagon D, Zou C, Lee W.
\newblock Modeling botnet propagation using time zones.
\newblock In: Annual Network \& Distributed System Security Symposium; 2006. .

\end{thebibliography}

\end{document}